# Optimal trade strategy of a regional economy by food exports

Madoka Okimoto [†]

University of Shizuoka

E-mail : [†] okimoto@u-shizuoka-ken.ac.jp

March 6, 2020

**Abstract**

This paper examines the export promotion of processed foods by a regional economy and regional vitalisation policy. We employ Bertrand models that contain a major home producer and a home producer in a local area. In our model, growth in the profit of one producer does not result in an increase in the profit of the other, despite strategic complements. We show that the profit of the producer in the local area decreases because of the deterioration of a location condition, and its profit increases through the reinforcement of the administrative guidance. Furthermore, when the inefficiency of the location worsens, the local government should optimally decrease the level of administrative guidance. Hence, the local government should strategically eliminate this inefficiency to maintain a sufficient effect of administrative guidance.

# 1. Introduction

Discussions have been increasing on the decline of regional economies and disparities between urban areas and regional areas in many countries, for example, as mentioned in Breinlich et al. (2013), the deterioration of Detroit, disparity between the northern and southern areas in the United Kingdom and Italy, overconcentration of the population in metropolitan areas in Japan, disparity between coastal areas and inland areas in China, and regional disparities in India and Brazil. Although the decline of regional economies shall also be regarded as an adjustment or natural consequence of rational accumulation in the field of economics, the restructuring or developing of regional economies is an issue for policymakers.

For the vitalisation of regional economies, producing and exporting food products from regional areas is considered an anticipated method because food is a necessity, and the interest in food safety and quality and health awareness is increasing worldwide. Walker(2016) argues that Detroit uses urban food production as an environmental justification and that Vancouver is attempting to win its international reputation as a cosmopolitan green city by using urban agriculture. Additionally, foods that attract vegetarians and vegans are receiving increasing attention in the market. Moreover, the cost of international transportation has decreased, tariffs on products have been removed in the long run, and intra-industry trade has recently increased. These reasons are why the discussion of how to derive a further benefit from foreign food markets is a critical research topic. For instance, the Norwegian aquaculture industry successfully exports farmed salmon. Liu et al. (2011) explain that Norwegian salmon aquaculture is governed by the public sector and that the purpose of the policy is developed from regional vitalisation to active advancements overseas.

We consider these matters in our attempt to provide theoretical insights into how producers and governments in local areas vitalise regional economies through the production and export of food and agricultural processed products. In the field of the theory of international trade, oligopoly models have frequently been adopted to study export strategies. Brander and Spencer(1985) provide a typical example of models of strategic trade policy and suppose that the producers compete in a Cournot fashion. With respect to Brander and Spencer(1985) and allied Cournot models, an export subsidy shifts the profit of the foreign producer towards the home producer and can improve the welfare of the home country. Contrary to those results in Cournot competition, Eaton and Grossman(1986) adopt the Bertrand paradigm and show that an export tax improves the home producer's profit and national welfare, instead of the export subsidy.

However, the problem is that an export subsidy or an export tax is inconsistent with the spirit of the World Trade Organization (WTO) and difficult to implement these days. For example, restraining export subsidies and policies that have equal effects in the agriculture sector have become a target in the agreement reached at the "Bali Package" in December 2013 (WTO, 2013), and nowadays, WTO members are moving forward with further regulation of fisheries subsidies (WTO, 2019). Hence, this paper models administrative guidance by the local government as a realistic export policy that does not contradict the spirit of the WTO, analyses the properties of the optimal policy, and discusses the impact of the policy on regional economies.

For that purpose, we develop oligopoly models in which two food producers, a producer located in an urban area, and a producer located in a local area export food products of the same type to a foreign country. For the producer located in the local area, the inefficiency due to its decentralised location is a major operational disadvantage, for example, the outflow of highly skilled human resources to urban areas. Accordingly, the producer in the local area requires a peculiar advantage to compete with other producers and increases corporate income. Thus, we assume that the producer in the local area that has such inefficiency can overcome it by adding new value to its food, and competes with the producer in the urban area, which denotes that both types of foods are imperfect substitutes.

In an oligopoly market where products are differentiated, the appropriate and realistic variable as the producers' strategy is considered the price rather than the output; therefore, we examine a third market model of Bertrand competition, as in Eaton and Grossman(1986). Notably, Eaton and Grossman(1986) reveal that when the equilibrium prices increase by imposing an export tax and politically shifting the home producer's reaction curve, the profit of foreign producer also increases, and an increase in prices leads to a decrease in demand for the home product and shrinkage in consumer surplus.

This paper develops non-specific and specific models to provide results with high robustness and concrete results. Our models differ from Eaton and Grossman(1986) as follows: the two producers in competition with each other are located in the same home country, and the adopted policy is not an export subsidy or an export tax but the administrative guidance by the local government, which reduces the production cost and improves the export strategies of the producer in the local area.

In our models, the location becoming inefficient and improvement in added value increase the production cost for the producer. Hence, from the consequences of Eaton and Grossman(1986), we deduce that such factors act the same as an export tax; consequently, the prices increase and both producers gain from those changes, because of strategic complements. Regarding the effects of the administrative guidance, we presume that a reduction in the production cost acts the same as an export subsidy, causing a decrease in prices and profit shrinkage while both producers gain from the improvement of the export strategies. However, these deductions do not necessarily correspond to our results.

First, the analysis of our models indicates that the deterioration in the location condition of the local area causes the price of food produced in any area to increase. By specifying the functions of the model, we show that the demand for food produced in the local area and consumer surplus shrink with this price hike, which corresponds to the results of Eaton and Grossman(1986). Second, a cost increase due to the inefficient location reduces the profit of the producer in the local area in our specific model, which is contrary to the result of Eaton and Grossman(1986). Third, although the effects of the improvement of added value and the strengthening of the administrative guidance on the prices depend on the structures of the demand and cost and are basically indeterminate, the specific model clarifies that the producer in the local area gains from the strengthening of such policy intervention.

Further, the specific model shows that because of the location becoming inefficient, the effect of increasing the profit by the policy shrinks, and as a consequence, the optimal level of the policy falls to balance the cost for the local government with the profit in the local area. This result leads us to presume that implementation of the administrative guidance at a certain degree of cost should be accompanied by the elimination of the inefficiency of the location; otherwise, this policy intervention would have failed, simply wasted cost, and in the long term, it decayed and vanished.

In addition to the aforementioned studies, this paper is related to the following theoretical literature of strategic trade policy or price competition: Youssef and Abderrazak(2009), Crespi and Marette(2001), Bastos et al. (2013), Okimoto(2015), Qui(1994), and Miller and Pazgal(2005). Blecha and Leitner(2014) and Horst et al. (2017) have conducted surveys on urban agriculture. Asche et al. (2019) explore food price in a gravity model.

The remainder of the paper is organised as follows. Section 2 introduces the theoretical models of food demand and production in the Bertrand paradigm. Section 3 presents the reaction functions of producers, and Section 4 analyses the properties of the Bertrand Nash equilibrium. Section 5 develops the game of the first stage, in which the local government sets the optimal administrative guidance for the vitalisation of regional economies and discusses the implication of the policy intervention. Section 6 concludes.

## 2. Models of food exports from the local area

We consider two food producers that operate in the same home country: a major food producer in an urban area (producer U) and a small or medium-sized producer located in a local area (producer L). Both produce and export food products of the same type to country X, in which many consumers exist. Producer L is inefficient due to its location, which leads to the disadvantage, for example, difficulty securing human resources, with $\bar{c}^L$ as the unit cost caused by the inefficiency. Producer L also has the advantage of having the technology particular to the local area, for instance, traditional or modern and advanced technology.

Such a technique of producer L for adding new value to its food (food L) can differentiate food L from the food of producer U (food U), enabling producer L to develop foods favoured by consumers in country X. We let $R$ be the added value per unit. However, the added value is no guarantee of success. Due to insufficient information on the sense of taste and the extent of the health consciousness in country X, the developed food L may not sell. $P(G)$ is the probability that food L suits the tastes of consumers in country X; $G$ denotes the level of administrative guidance of the local government to make producer L operate more efficiently; hence, $P'(G) > 0$.

### 2.1. A model of a non-specific framework

Considering the aforementioned, we define the non-specific demand function of food $j$ in country X as $X^j(p^U, p^L; R, G)$, where (i) $\frac{\partial X^j}{\partial p^j} < 0$, $\frac{\partial X^j}{\partial p^k} > 0$, $j, k = U, L$, and $j \neq k$, and (ii) $\frac{\partial X^U}{\partial t} < 0$, $\frac{\partial X^L}{\partial t} > 0$, and $t = R, G$. $p^j$ ($p^k$) denotes the price of food $j$ ($k$). Under the Bertrand competition, producer $j$ chooses the price of food $j$ to maximise its profit:

$$\max_{p^j} \pi^j = (p^j - c^j)X^j \qquad j = U, L,$$

where $\pi^j$ denotes the profit and $c^j$ is the unit cost of producer $j$. We assume that $c^U$ is positive and constant. $c^L$ is the function that satisfies $\frac{\partial c^L}{\partial \bar{c}^L} > 0$, $\frac{\partial c^L}{\partial R} > 0$, and $\frac{\partial c^L}{\partial G} < 0$.

### 2.2. A model of a specific framework

Next, we specify the functional form and obtain a demand function that is easy to handle. First, we let $\theta_i$ be the extent of consumer $i$'s preference for the added value, $R$.

According to $\theta_i$, consumers are uniformly distributed over [0,1]. When the added value suits the tastes of consumers, the higher $\theta_i$, the more consumer $i$ prefers the added value. By contrast, when they are not concerned with the added value, $\theta_i = 0$ holds for all consumers. Because food U has no added value ($R = 0$), the utility of consumer $i$ obtained from one unit of food is summarised as follows:

$$u_i = \begin{cases} \sqrt{q} + R\theta_i & \text{if } R > 0 \text{ and } \theta_i \neq 0 \\ \sqrt{q} & \text{if } R = 0 \text{ and/or } \theta_i = 0 \end{cases},$$

where $q$ is the basic utility of food that is common to two types of food; the concavity of the basic utility function, $\sqrt{q}$, shows the risk aversion of each consumer.

Now, we can define the utility of consumer $i$ obtained from one unit of food $j$:

$$u_i^U = \sqrt{q},$$
$$u_i^L = P(G)\left(\sqrt{q} + R\theta_i\right) + \left(1 - P(G)\right)\sqrt{q}.$$

Given our utility specification, consumer surpluses by purchasing food U and food L for consumer $i$ with $\theta_i$ are, respectively,

$$CS_i^U = \sqrt{q} - p^U,$$
$$CS_i^L = P(G)\left(\sqrt{q} + R\theta_i\right) + \left(1 - P(G)\right)\sqrt{q} - p^L.$$

Comparing two levels of consumer surplus, consumer $i$ chooses food U or food L and purchases at most one food $j$. However, consumer $i$ has an incentive to purchase food $j$ only when he/she obtains a non-negative consumer surplus by purchasing food $j$ ($CS_i^j \geq 0$). For simplicity, we make

**Condition 1** $CS_i^U > 0 \Leftrightarrow \sqrt{q} > p^U$.

Condition 1 implies that all consumers can obtain a positive consumer surplus from food U. With respect to food L, $CS_i^L \geq 0 \Leftrightarrow \theta_i \geq \frac{p^L - \sqrt{q}}{P(G)R}$ leads to the threshold that the incentive to purchase food L vanishes as $\theta_i^* \equiv \frac{p^L - \sqrt{q}}{P(G)R}$. Here, we set

**Condition 2**  $\theta_i^* > 0 \Leftrightarrow p^L > \sqrt{q}$.

Condition 2 implies that some consumers have no incentive to purchase food L.

Conditions 1 and 2 yield $p^L > p^U$. $CS_i^U \gtreqless CS_i^L \Leftrightarrow \theta_i \lesseqgtr \frac{p^L - p^U}{P(G)R}$ leads to the threshold of demand as $\theta_i^{**} \equiv \frac{p^L - p^U}{P(G)R}$. Therefore, we find that $\theta_i^{**} > \theta_i^* > 0$. Additionally, we make

**Condition 3**  $1 > \theta_i^{**} \Leftrightarrow P(G)R > p^L - p^U$.

Condition 3 implies that some consumers purchase food L. Notably, condition 3 is not necessarily true of the real economy but is necessary to analyse the market where both types of food are consumed. This specified preference structure leads to the following demand functions:

$$X^U = \theta_i^{**} = \frac{p^L - p^U}{P(G)R},$$

$$X^L = 1 - \theta_i^{**} = 1 - \frac{p^L - p^U}{P(G)R}.$$

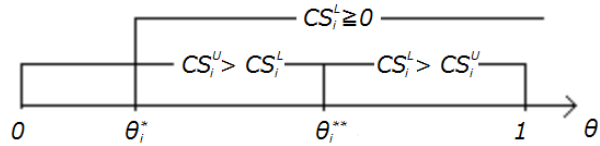

Figure 1  Threshold of demand over the preference for the added value.

We define the unit costs of producer U and producer L as $c^U = \bar{c}$ and $c^L = \bar{c} + \bar{c}^L + \alpha(R, G)$, respectively. $\bar{c}$ is the basic unit cost. $\alpha(R, G)$ is the additional unit cost which provides the added value where $\frac{\partial \alpha}{\partial R} > 0$ and $\frac{\partial \alpha}{\partial G} < 0$. In the Bertrand paradigm, producers choose prices to maximise their profit:

$$\max_{p^U} \pi^U = (p^U - c^U)X^U = (p^U - \bar{c})\frac{p^L - p^U}{P(G)R},$$

$$\max_{p^L} \pi^L = (p^L - c^L)X^L = [p^L - (\bar{c} + \bar{c}^L + \alpha(R,G))]\left(1 - \frac{p^L - p^U}{P(G)R}\right).$$

## 3. Stability condition and iso-profit curves under Bertrand competition

To indicate variables of the Bertrand Nash equilibrium, the subscript "E" is used.

### 3.1. Non-specific model

In the non-specific model, the first-order condition of producer $j$ is

$$\frac{\partial \pi^j}{\partial p^j} = \frac{\partial X^j}{\partial p^j}(p^j - c^j) + X^j = 0 \qquad j = U, L.$$

Define $J$, $\boldsymbol{P}$, $\boldsymbol{C}$, $\boldsymbol{R}$, and $\boldsymbol{G}$ as $J \equiv \begin{bmatrix} \frac{\partial^2 \pi^U}{\partial p^{U^2}} & \frac{\partial^2 \pi^U}{\partial p^U \partial p^L} \\ \frac{\partial^2 \pi^L}{\partial p^L \partial p^U} & \frac{\partial^2 \pi^L}{\partial p^{L^2}} \end{bmatrix}$, $\boldsymbol{P} \equiv {}^t[p^U \quad p^L]$, $\boldsymbol{C} \equiv {}^t\left[\frac{\partial^2 \pi^U}{\partial p^U \partial \bar{c}^L} \quad \frac{\partial^2 \pi^L}{\partial p^L \partial \bar{c}^L}\right]$,

$\boldsymbol{R} \equiv {}^t\left[\frac{\partial^2 \pi^U}{\partial p^U \partial R} \quad \frac{\partial^2 \pi^L}{\partial p^L \partial R}\right]$, and $\boldsymbol{G} \equiv {}^t\left[\frac{\partial^2 \pi^U}{\partial p^U \partial G} \quad \frac{\partial^2 \pi^L}{\partial p^L \partial G}\right]$. Differentiate the first-order conditions and use these matrix and vectors to provide the properties of food prices in the equilibrium:

$$J d\boldsymbol{P} = -\boldsymbol{C} d\bar{c}^L - \boldsymbol{R} dR - \boldsymbol{G} dG. \tag{1}$$

We assume the second-order conditions are satisfied, that is, $\frac{\partial^2 \pi^U}{\partial p^{U^2}} < 0$ and $\frac{\partial^2 \pi^L}{\partial p^{L^2}} < 0$.

Next, we assume that the food prices adjust with time according to the conditions, $\dot{p}^U = k^U(p^{UE} - p^U)$ and $\dot{p}^L = k^L(p^{LE} - p^L)$; a "dot" represents the change in a variable according to time. We let $Z$ be the variable which denotes a sort of distance from the equilibrium and define $2Z = k^U(p^{UE} - p^U)^2 + k^L(p^{LE} - p^L)^2$. Under the adjustment processes according to time and the reaction functions, the stability condition of the equilibrium is equal to the condition which makes $Z$ become a Liapunov function under Bertrand competition:[1]

---

[1] Based on the ideas of global stability using a Liapunov function, Wong(1995) shows the

**Condition 4 (Stability condition)** $\quad 0 < a < -b \quad$ and $\quad 0 < c < -d,$

where $a \equiv \frac{\partial X^U}{\partial p^L} + \frac{\partial^2 X^U}{\partial p^U \partial p^L}(p^U - c^U)$, $b \equiv \frac{\partial X^U}{\partial p^U} + \frac{\partial^2 X^U}{\partial p^{U^2}}(p^U - c^U)$, $c \equiv \frac{\partial X^L}{\partial p^U} + \frac{\partial^2 X^L}{\partial p^L \partial p^U}(p^L - c^L)$,

and $d \equiv \frac{\partial X^L}{\partial p^L} + \frac{\partial^2 X^L}{\partial p^{L^2}}(p^L - c^L)$ (see Appendix A). Using condition 4 and $\frac{\partial X^j}{\partial p^j} < 0$, we find that $|J| = \left(b + \frac{\partial X^U}{\partial p^U}\right)\left(d + \frac{\partial X^L}{\partial p^L}\right) - ac > 0$.

The relationship of the slopes of reaction functions is $\left.\frac{dp^U}{dp^L}\right|_L - \left.\frac{dp^U}{dp^L}\right|_U = \frac{\left(b + \frac{\partial X^U}{\partial p^U}\right)\left(d + \frac{\partial X^L}{\partial p^L}\right) - ac}{-c\left(b + \frac{\partial X^U}{\partial p^U}\right)} > 0$, with $\left.\frac{dp^U}{dp^L}\right|_U = \frac{-a}{b + \frac{\partial X^U}{\partial p^U}} > 0 \quad$ and $\quad \left.\frac{dp^U}{dp^L}\right|_L = \frac{-\left(d + \frac{\partial X^L}{\partial p^L}\right)}{c} > 0 \quad$ being the

slopes of the reaction functions of producer U and producer L, respectively. With respect to producer U, the total differential, $d\pi^U = 0$, leads to the slope of its iso-profit curve as

$\frac{dp^L}{dp^U} = -\frac{X^U + (p^U - c^U)\frac{\partial X^U}{\partial p^U}}{(p^U - c^U)\frac{\partial X^U}{\partial p^L}}$ and the total differential of $\pi^U$ where $dp^U = 0$ yields $\frac{d\pi^U}{dp^L} = (p^U - c^U)\frac{\partial X^U}{\partial p^L} > 0$. Similarly, the properties of the iso-profit curve of producer L are obtained as $\frac{dp^U}{dp^L} = -\frac{X^L + (p^L - c^L)\frac{\partial X^L}{\partial p^L}}{(p^L - c^L)\frac{\partial X^L}{\partial p^U}}$ and $\frac{d\pi^L}{dp^U} = (p^L - c^L)\frac{\partial X^L}{\partial p^U} > 0$. For simplicity, we assume

$\frac{d^2 p^L}{dp^{U^2}} > 0$ and $\frac{d^2 p^U}{dp^{L^2}} > 0$, based on the ordinary Bertrand model(Appendix B). Under these assumptions, the iso-profit curves of producer U are convex towards the left side, and the profit becomes higher as the curve is located on the right side more in the figure with the set $p^U$ as the ordinate. By contrast, the iso-profit curves of producer L are convex downward, and the profit becomes higher as the curve is more upward in that figure.

---

conditions for stability of the Bertrand Nash equilibrium under the non-specified demand functions as given by Eaton and Grossman(1986). We use the stability conditions in Wong(1995); notably, these two stability conditions, respectively, correspond to the cases of strategic complements and strategic substitutes; our stability condition corresponds to the former.

## 3.2. Specific model

In the specific model, the first-order conditions are

$$\frac{\partial \pi^U}{\partial p^U} = 0 \Leftrightarrow p^U = \frac{1}{2}(p^L + \bar{c}), \tag{2a}$$

$$\frac{\partial \pi^L}{\partial p^L} = 0 \Leftrightarrow p^L = \frac{1}{2}\left[P(G)R + p^U + \left(\bar{c} + \bar{c}^L + \alpha(R, G)\right)\right]. \tag{2b}$$

These conditions lead to equilibrium prices, $p^{UE} = \frac{1}{3}\left(P(G)R + \bar{c}^L + \alpha(R, G)\right) + \bar{c}$ and $p^{LE} = \frac{2}{3}\left(P(G)R + \bar{c}^L + \alpha(R, G)\right) + \bar{c}$. In this case, we obtain $\frac{\partial^2 \pi^U}{\partial p^{U2}} < 0$ and $\frac{\partial^2 \pi^L}{\partial p^{L2}} < 0$. $\left.\frac{dp^U}{dp^L}\right|_L = 2 > \frac{1}{2} = \left.\frac{dp^U}{dp^L}\right|_U$ enables the stability condition according to the reaction functions to hold. Additionally, the adjustment process according to time always remains stable (Appendix A), and with the set $p^U$ as the ordinate, the iso-profit curves of producer U are convex towards the left ($\frac{d^2 p^L}{dp^{U2}} = \frac{2P(G)R\pi^U}{(p^U - \bar{c})^3} > 0$), and the profit becomes higher as the curve is more right. Similarly, the iso-profit curves of producer L are convex downwards ($\frac{d^2 p^U}{dp^{L2}} = \frac{2P(G)R\pi^L}{(p^L - c^L)^3} > 0$), and the profit becomes higher as the curve is more upwards.

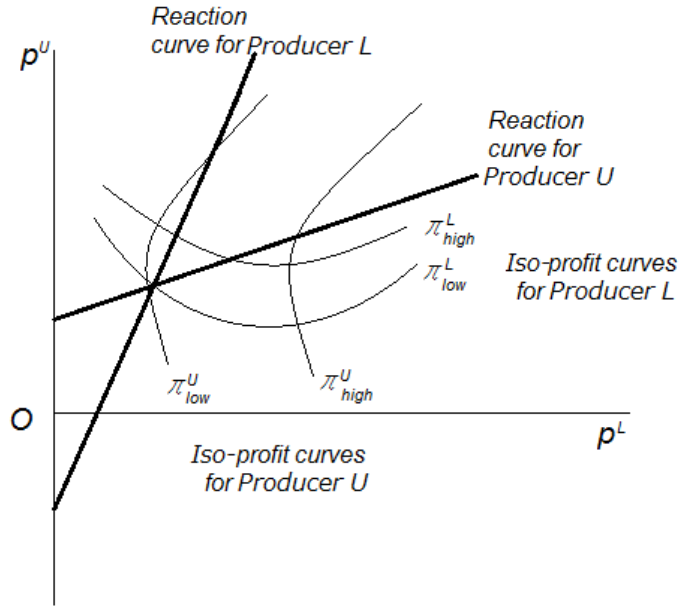

Figure 2 Reaction curves and iso-profit curves of the ordinary Bertrand model.

## 4. Comparative statics of the second stage

### 4.1. Equilibrium prices

First, the results with respect to $\bar{c}^L$ obtained by Eq.(1) in the non-specific model are

$$\frac{\partial p^{LE}}{\partial \bar{c}^L} = \frac{1}{|J|}\left(b + \frac{\partial X^U}{\partial p^U}\right)\frac{\partial X^L}{\partial p^L}\frac{\partial c^L}{\partial \bar{c}^L} > \frac{\partial p^{UE}}{\partial \bar{c}^L} = -\frac{1}{|J|}a\frac{\partial X^L}{\partial p^L}\frac{\partial c^L}{\partial \bar{c}^L} > 0$$

because $-b > a$. Eqs. (2) in the specific model also leads to $\frac{\partial p^{LE}}{\partial \bar{c}^L} > \frac{\partial p^{UE}}{\partial \bar{c}^L} > 0$. Hence, we find

**Proposition 1.** *An increase in the price of food produced in the local area is greater than an increase in the price of food produced in the urban area, as the inefficiency of the location in a local area worsens.*

Clearly, the propagation of an increase in $p^L$ to $p^U$ occurs through a positive indirect effect behind Proposition 1 because the direct effects common to both models are $\left.\frac{dp^U}{d\bar{c}^L}\right|_U = 0$ and $\left.\frac{dp^L}{d\bar{c}^L}\right|_L > 0$. This finding implies that the inconvenient location of producers in the local area causes a nationwide hike in the price of food because of strategic complements.

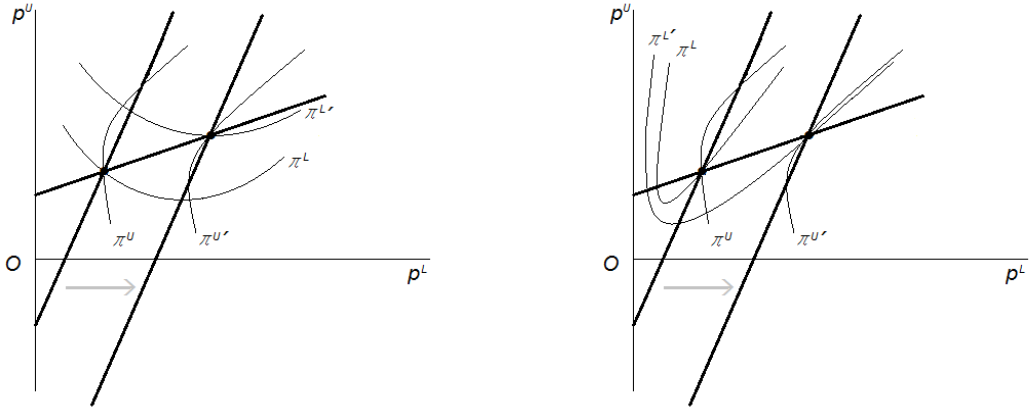

3-1 An ordinary case    3-2 The specific model

Figure 3  Change in the inefficiency of the location.

Second, the results with respect to $R$ and $G$ in the non-specific model are indeterminate and summarised as follows:

$$|J|\frac{\partial p^{UE}}{\partial t} \approx -\left[\frac{\partial X^U}{\partial t}\left(d + \frac{\partial X^L}{\partial p^L}\right) - a\left(-\frac{\partial X^L}{\partial p^L}\frac{\partial c^L}{\partial t} + \frac{\partial X^L}{\partial t}\right)\right] \quad t = R, G,$$

$$|J|\frac{\partial p^{LE}}{\partial t} \approx -\left(b + \frac{\partial X^U}{\partial p^U}\right)\left(-\frac{\partial X^L}{\partial p^L}\frac{\partial c^L}{\partial t} + \frac{\partial X^L}{\partial t}\right) + \frac{\partial X^U}{\partial t}c \quad t = R, G,$$

where these approximations assume that $\frac{\partial^2 X^U}{\partial p^U \partial t}$ and $\frac{\partial^2 X^L}{\partial p^L \partial t}$ are sufficiently small (Appendix C). Therefore, we obtain

$$\text{sign}\left(\frac{\partial p^{UE}}{\partial t}\right) \approx \text{sign}\left[-\left|\frac{\partial^2 \pi^L}{\partial p^{L2}}\right|\left|\frac{\partial X^U}{\partial t}\right| + \frac{\partial^2 \pi^U}{\partial p^U \partial p^L}\left(\left|\frac{\partial X^L}{\partial p^L}\right|\frac{\partial c^L}{\partial t} + \frac{\partial X^L}{\partial t}\right)\right], \quad (3a)$$

$$\text{sign}\left(\frac{\partial p^{LE}}{\partial t}\right) \approx \text{sign}\left[\left|\frac{\partial^2 \pi^U}{\partial p^{U2}}\right|\left(\left|\frac{\partial X^L}{\partial p^L}\right|\frac{\partial c^L}{\partial t} + \frac{\partial X^L}{\partial t}\right) - \frac{\partial^2 \pi^L}{\partial p^L \partial p^U}\left|\frac{\partial X^U}{\partial t}\right|\right]. \quad (3b)$$

We assume that $\left|\frac{\partial^2 \pi^U}{\partial p^{U2}}\right|$, $\left|\frac{\partial^2 \pi^L}{\partial p^{L2}}\right|$, $\frac{\partial^2 \pi^U}{\partial p^U \partial p^L}$ and $\frac{\partial^2 \pi^L}{\partial p^L \partial p^U}$ [2] are also sufficiently small.

Utilising the assumption that $\frac{\partial X^U}{\partial R} < 0$ and $\frac{\partial X^L}{\partial R} > 0$, we interpret Eqs. (3) when $t = R$; as far as influences such that $\frac{\partial X^L}{\partial R} > 0$, $\frac{\partial c^L}{\partial R} > 0$, and $\frac{\partial X^L}{\partial p^L} < 0$ on producer L are (respectively, influence such that $\frac{\partial X^U}{\partial R} < 0$ on producer U is) so large, $\frac{\partial p^{UE}}{\partial R} > 0$ and $\frac{\partial p^{LE}}{\partial R} > 0$ (respectively, $\frac{\partial p^{UE}}{\partial R} < 0$ and $\frac{\partial p^{LE}}{\partial R} < 0$) in the non-specific model. By contrast, $\frac{\partial p^{LE}}{\partial R} > \frac{\partial p^{UE}}{\partial R} > 0$ is obtained in the specific model. These lead to

**Proposition 2.** *An improvement in the added value of food produced in the local area causes food price increases if increases in the demand for and production cost for the food produced in the local area, and gross self-substitution effect of that are large.*

Eq. (1) shows that we have $\left.\frac{dp^U}{dR}\right|_U \approx \frac{\frac{\partial X^U}{\partial R}}{-\frac{\partial^2 \pi^U}{\partial p^{U2}}} < 0$ and $\left.\frac{dp^L}{dR}\right|_L \approx \frac{-\frac{\partial X^L}{\partial p^L}\frac{\partial c^L}{\partial R} + \frac{\partial X^L}{\partial R}}{-\frac{\partial^2 \pi^L}{\partial p^{L2}}} > 0$ as the

---

[2] $0 < a$ and $0 < c$ are assumed in condition 4, where $a \equiv \frac{\partial^2 \pi^U}{\partial p^U \partial p^L}$ and $c \equiv \frac{\partial^2 \pi^L}{\partial p^L \partial p^U}$.

approximate direct effects in the non-specific model (Figure 4). Likewise, Eqs. (2) indicate that $\frac{dp^U}{dR}\big|_U = 0$ and $\frac{dp^L}{dR}\big|_L > 0$ are the direct effects in the specific model as in Figure 3. We consider these matters and presume that whether the equilibrium prices increase depends on the magnitude correlation between such a $p^L$ increase and $p^U$ decrease or invariance as the direct effect, because the strategies of the producers are complementary.[3]

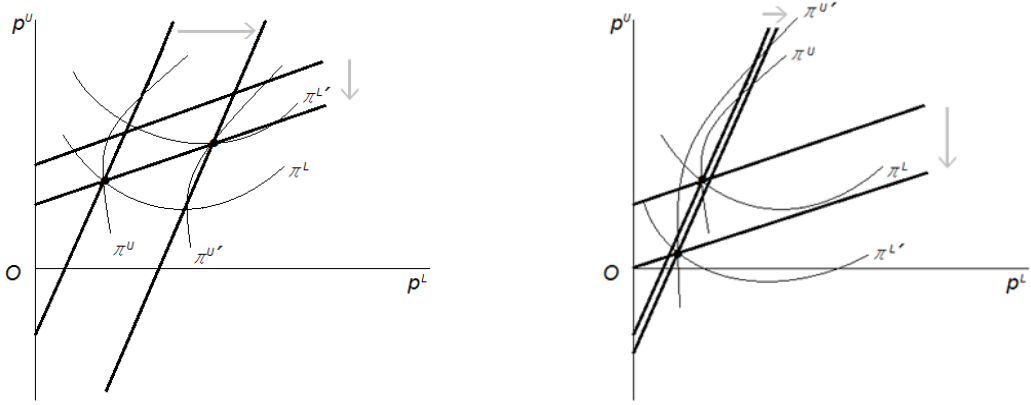

4-1   High reaction of Producer L     4-2   High reaction of Producer U

($\frac{\partial p^{UE}}{\partial R} > 0$ and $\frac{\partial p^{LE}}{\partial R} > 0$)      ($\frac{\partial p^{UE}}{\partial R} < 0$ and $\frac{\partial p^{LE}}{\partial R} < 0$)

Figure 4   Change in the added value in the non-specific model.

Here, we utilize the assumption that $\frac{\partial X^U}{\partial G} < 0$ and $\frac{\partial X^L}{\partial G} > 0$ to interpret Eqs. (3) when $t = G$; as far as $\frac{\partial X^U}{\partial G} < 0$, $\frac{\partial c^L}{\partial G} < 0$, and $\frac{\partial X^L}{\partial p^L} < 0$ are (respectively, $\frac{\partial X^L}{\partial G} > 0$ is) sufficiently large, that $\frac{\partial p^{UE}}{\partial G} < 0$ and $\frac{\partial p^{LE}}{\partial G} < 0$ (respectively, $\frac{\partial p^{UE}}{\partial G} > 0$ and $\frac{\partial p^{LE}}{\partial G} > 0$). We obtain a similar result, namely, $\text{sign}\left(\frac{\partial p^{UE}}{\partial G}\right) = \text{sign}\left(\frac{\partial p^{LE}}{\partial G}\right) = \text{sign}\left(P'(G)R + \frac{\partial \alpha}{\partial G}\right)$, in the specific model. These provide

---

[3] Since the conditions that lead $\frac{\partial p^{UE}}{\partial R} > 0$ and $\frac{\partial p^{LE}}{\partial R} > 0$ are the same quality, two prices tend to move in a parallel direction (Figure 4).

**Proposition 3.** *An increase in the level of administrative guidance causes a decrease in food prices if a decrease in the demand for the food produced in the urban area and effect of the cost reduction and gross self-substitution effect of the food produced in the local area are large.*

Compared with Eqs. (3) when $t = R$, the factor of the price increase is reduced in Eqs. (3) when $t = G$: the administrative guidance relatively hardly causes a food price hike. This difference results from the indefiniteness of the direct effect of $G$ on $p^L$ because of the cost reduction effect ($\frac{\partial c^L}{\partial G} < 0$ and $\frac{\partial \alpha}{\partial G} < 0$), in contrast to the positive direct effect of $R$ on $p^L$: $\frac{dp^L}{dG}\Big|_L \approx \frac{-\frac{\partial x^L}{\partial p^L}\frac{\partial c^L}{\partial G}+\frac{\partial x^L}{\partial G}}{-\frac{\partial^2 \pi^L}{\partial p^{L2}}}$ in the non-specific case and $\frac{dp^L}{dG}\Big|_L = \frac{1}{2}\left(P'(G)R + \frac{\partial \alpha}{\partial G}\right)$ in the specific case. Thus, a sufficient cost reduction through the promotion of local industry leads to an avoidance of a price hike. Figure 4 analogises the case of $\frac{dp^L}{dG}\Big|_L > 0$, and Figure 5 illustrates the case of $\frac{dp^L}{dG}\Big|_L < 0$.[4] Provided that $\frac{dp^U}{dG}\Big|_U \approx \frac{\frac{\partial x^U}{\partial G}}{-\frac{\partial^2 \pi^U}{\partial p^{U2}}} < 0$ in the non-specific case and that $\frac{dp^U}{dG}\Big|_U = 0$ in the specific case have the same quality as those of $R$.

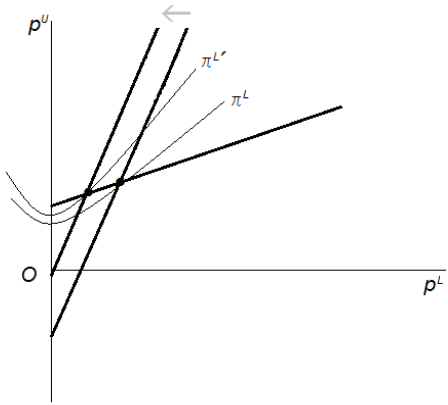
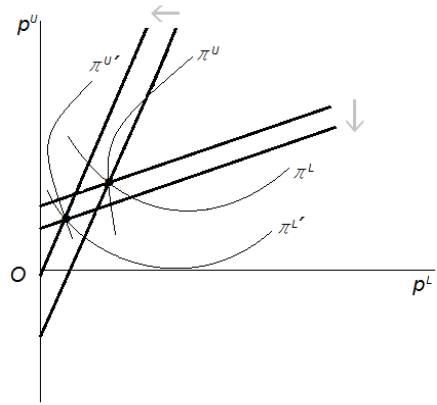

5-1  The specific model    5-2  The non-specific model
Figure 5   Change in administrative guidance: In the case of price decreases.

---

[4] The iso-profit curves in Figure 5-2 are standard but not necessarily true.

## 4.2. Equilibrium demands, consumer surpluses, and profit levels

The non-specific model is of little use to disclose the properties of demands, consumer surpluses, and profit levels in the second stage equilibrium. Therefore, we focus on the specific model to analyse those properties in this section.

With respect to $\bar{c}^L$, we obtain $\frac{\partial X^{UE}}{\partial \bar{c}^L} > 0$, $\frac{\partial X^{LE}}{\partial \bar{c}^L} < 0$, $\frac{\partial CS_i^{UE}}{\partial \bar{c}^L} < 0$, and $\frac{\partial CS_i^{LE}}{\partial \bar{c}^L} < 0$: a part of demands flow out of food L and into food U, and the foreign consumer surplus shrinks. Although the ordinary Bertrand paradigm (Figure 3-1) shows that increasing the production cost leads to the increase in equilibrium profits of both producers, our specific model shows that the equilibrium profit of producer U increases ($\frac{\partial \pi^{UE}}{\partial \bar{c}^L} > 0$) and that of producer L decreases ($\frac{\partial \pi^{LE}}{\partial \bar{c}^L} < 0$) (Figure 3-2).[5]

By contrast, the information is insufficient on how the second stage equilibrium is affected by $R$: the signs of $\frac{\partial X^{UE}}{\partial R}$, $\frac{\partial X^{LE}}{\partial R}$, $\frac{\partial CS_i^{LE}}{\partial R}$, $\frac{\partial \pi^{UE}}{\partial R}$, and $\frac{\partial \pi^{LE}}{\partial R}$ are indeterminate, and $\frac{\partial CS_i^{UE}}{\partial R} < 0$, despite the specific model. Regarding $G$, we obtain $\frac{\partial X^{UE}}{\partial G} < 0$ and $\frac{\partial X^{LE}}{\partial G} > 0$ in that model. Notably, $\frac{\partial \pi^{LE}}{\partial G} > 0$ is obtained even though the signs of $\frac{\partial p^{UE}}{\partial G}$ and $\frac{\partial p^{LE}}{\partial G}$ are ambiguous.[6] Unfortunately, the sign of $\frac{\partial CS_i^{UE}}{\partial G}$, $\frac{\partial CS_i^{LE}}{\partial G}$, and $\frac{\partial \pi^{UE}}{\partial G}$ is indeterminate.

## 5. Local welfare

In the first stage, the local government chooses the level of $G$ to maximise local welfare, which is defined as $W \equiv \pi^{LE} - \beta(G)$. $\beta(G)$ denotes the administrative cost and $\beta'(G) > 0$. As it is difficult to provide a detailed analysis of the first stage by using the

---

[5] $\frac{\partial \pi^{UE}}{\partial \bar{c}^L} = \frac{p^{LE} - \bar{c}}{3P(G)R} > 0$. The envelope theorem and $\frac{\partial \pi^L}{\partial p^L} = 0 \Leftrightarrow P(G)R + p^U = 2p^L - c^L$ clarify $\frac{\partial \pi^{LE}}{\partial \bar{c}^L} = \frac{1}{P(G)R}\left[\frac{4}{3}p^{LE} - \frac{1}{3}c^L - (P(G)R + p^{UE})\right] = \frac{-2(p^{LE} - c^L)}{3P(G)R} < 0$.

[6] Obtained by partial differentiation:

$\frac{\partial \pi^{LE}}{\partial G} = P'(G)\frac{p^{LE} - c^L}{P(G)}\left(\frac{1}{3} + \frac{p^{LE} - p^{UE}}{P(G)R}\right) + \frac{\partial \alpha}{\partial G}\frac{1}{P(G)R}\left[\frac{1}{3}(p^{LE} - c^L) - P(G)R + (p^{LE} - p^{UE})\right]$ where the condition $\frac{\partial \pi^L}{\partial p^L} = 0 \Leftrightarrow p^U = 2p^L - c^L - P(G)R$ leads to $\left[\frac{1}{3}(p^{LE} - c^L) - P(G)R + (p^{LE} - p^{UE})\right] = -\frac{2}{3}(p^{LE} - c^L) < 0$.

non-specific model, hereafter, we use the specific model. Hence, the problem of the local government is

$$\max_{G} W = \pi^{LE} - \beta(G) = [p^{LE} - (\bar{c} + \bar{c}^L + \alpha(R,G))]\left(1 - \frac{p^{LE} - p^{UE}}{P(G)R}\right) - \beta(G).$$

The first-order condition is $\frac{\partial W}{\partial G} = 0 \Leftrightarrow \frac{\partial \pi^{LE}}{\partial G} = \beta'(G)$, which provides that the optimal administrative guidance balances the increasing profit and increasing administrative cost of the local area. The second-order condition is assumed to be satisfied as $\frac{\partial^2 W}{\partial G^2} \equiv \rho < 0$.

As the equilibrium profit of producer L can be written as $\pi^{LE} = \pi^L[p^{UE}(G), p^{LE}(G), G]$ and the envelope theorem holds ($\frac{\partial \pi^L}{\partial p^L} = 0$ under $p^L = p^{LE}$), $\frac{\partial \pi^{LE}}{\partial G} > 0$ can be decomposed into the direct effect and indirect effect:

$$\frac{\partial \pi^{LE}}{\partial G} = \frac{\partial \pi^L}{\partial p^U}\frac{\partial p^{UE}}{\partial G} + \frac{\partial \pi^L}{\partial G}$$
$$= \frac{p^{LE} - c^L}{P(G)R}\frac{\partial p^{UE}}{\partial G} + \left(1 - \frac{p^{LE} - p^{UE}}{P(G)R}\right)\left(-\frac{\partial \alpha}{\partial G}\right) + \frac{p^{LE} - c^L}{P(G)}\frac{p^{LE} - p^{UE}}{P(G)R}P'(G) > 0.$$

The aforementioned equation identifies two effects that contribute to the increase in $\pi^{LE}$: one is $P'(G)$ and the other is $\frac{\partial \alpha}{\partial G}$. However, the effect of $\frac{\partial p^{UE}}{\partial G}$ on $\pi^{LE}$ is indeterminate.

We totally differentiate the implicit function expression, $\frac{\partial \pi^{LE}}{\partial G} = \beta'(G)$, and reveal how the optimal administrative guidance, defined as $G^E$, depends on $\bar{c}^L$ as

$$\frac{\partial G^E}{\partial \bar{c}^L} = \frac{-1}{\rho}\frac{\partial}{\partial \bar{c}^L}\left(\frac{\partial \pi^{LE}}{\partial G} - \beta'(G)\right) = \frac{-1}{\rho}\frac{\partial^2 \pi^{LE}}{\partial G \partial \bar{c}^L}.$$

Because $\frac{\partial p^{UE}}{\partial G}$, $\frac{\partial \alpha}{\partial G}$ and $P'(G)$ do not depend on $\bar{c}^L$, $\frac{\partial^2 \pi^{LE}}{\partial G \partial \bar{c}^L}$ can be decomposed into the effects of $\frac{\partial p^{UE}}{\partial G}$, $\frac{\partial \alpha}{\partial G}$ and $P'(G)$ as

$$\begin{aligned}
\frac{\partial^2 \pi^{LE}}{\partial G \partial \bar{c}^L} &= \frac{\partial}{\partial \bar{c}^L}\left(\frac{\partial \pi^{LE}}{\partial G}\right) \\
&= \frac{\partial}{\partial \bar{c}^L}\left(\frac{p^{LE} - c^L}{P(G)R}\right)\frac{\partial p^{UE}}{\partial G} + \frac{\partial}{\partial \bar{c}^L}\left(1 - \frac{p^{LE} - p^{UE}}{P(G)R}\right)\left(-\frac{\partial \alpha}{\partial G}\right) \\
&\quad + \frac{\partial}{\partial \bar{c}^L}\left(\frac{p^{LE} - c^L}{P(G)}\frac{p^{LE} - p^{UE}}{P(G)R}\right)P'(G) \\
&= \frac{-1}{3P(G)R}\frac{\partial p^{UE}}{\partial G} + \frac{-1}{3P(G)R}\left(-\frac{\partial \alpha}{\partial G}\right) + \frac{(p^{UE} - c^L)}{3(P(G))^2 R}P'(G).
\end{aligned}$$

Namely, an increase in $\bar{c}^L$ mitigates the effects of a change in price ($\frac{\partial p^{UE}}{\partial G}$) and a cost cut ($-\frac{\partial \alpha}{\partial G} > 0$) on $\frac{\partial \pi^{LE}}{\partial G}$, and how an increase in $\bar{c}^L$ influences the effect of $P'(G)$ on $\frac{\partial \pi^{LE}}{\partial G}$ depends on the sign of $(p^{UE} - c^L)$ and is ambiguous. Furthermore, $\frac{\partial^2 \pi^{LE}}{\partial G \partial \bar{c}^L}$ reduces to

$$\frac{\partial^2 \pi^{LE}}{\partial G \partial \bar{c}^L} = \frac{1}{3P(G)R}\left[\frac{2}{3}\frac{\partial \alpha}{\partial G} + \frac{-P(G)R\frac{1}{3} + (p^{UE} - c^L)}{P(G)}P'(G)\right] < 0.[7]$$

$\frac{\partial^2 \pi^{LE}}{\partial G \partial \bar{c}^L} < 0$ means that as the location becomes inefficient, the additional profit caused by the policy intervention, $\frac{\partial \pi^{LE}}{\partial G}$, reduces. That reveals $\frac{\partial G^E}{\partial \bar{c}^L} = \frac{-1}{\rho}\frac{\partial^2 \pi^{LE}}{\partial G \partial \bar{c}^L} < 0$ and leads to

**Proposition 4.** *A deterioration in the location condition for the producer in the local area leads to a lower level of optimal administrative guidance.*

Because a profit increase by $G$ ($\frac{\partial \pi^{LE}}{\partial G} > 0$) shrinks because of a higher inefficiency of the location, and a cost increase for $G$ ($\beta'(G) > 0$) does not depend on this inefficiency, Proposition 4 clearly holds to sustain the optimal balance between marginal benefit and marginal cost($\frac{\partial \pi^{LE}}{\partial G} = \beta'(G)$). Proposition 4 is intuitively correct and results from $\frac{\partial^2 \pi^{LE}}{\partial G \partial \bar{c}^L} < 0$, which is also plausible even though we obtain it by specifying the model.

---

[7] Notably, $-P(G)R\frac{1}{3} + (p^{UE} - c^L) = -\frac{2}{3}(\bar{c}^L + \alpha(R,G)) < 0$ because $p^{UE} = \frac{1}{3}(P(G)R + \bar{c}^L + \alpha(R,G)) + \bar{c}$ and $c^L = \bar{c} + \bar{c}^L + \alpha(R,G)$.

In contrast with Proposition 4, the effect of the added value on the optimal administrative cost is ambiguous, namely, $\text{sign}\left(\frac{\partial G^E}{\partial R}\right) = \text{sign}\left(\frac{\partial^2 \pi^{LE}}{\partial G \partial R}\right)$, and the sign of $\frac{\partial^2 \pi^{LE}}{\partial G \partial R}$ is indeterminate. The added value yields both a benefit and a production cost; thus, which of the two effects outweighs the other is indeterminate.

## 6. Conclusion

This paper describes export promotion of agricultural processed products with a commodity added value or high-quality foods by a regional economy which has a disadvantage of decentralised location. We theoretically analyse the impact of the inefficiency due to the location, the added value to foods, and the administrative guidance by the local government on the regional economy and examine regional vitalisation policies.

Based on Eaton and Grossman(1986), we develop Bertrand-competition models with two home producers: a major producer in an urban area and a small or medium-sized producer in a local area that produce and export two foods that are imperfect substitutes to a particular market outside, and price competition occurs in that market. In our models, the producer in the local area has a peculiar technology and adds new value to its food to overcome its inefficient location. Such a model has shown that as the inefficiency due to the location in the local area worsens, the prices of food produced in the local and urban areas increase, a part of demands flows out of food produced in the local area into food produced in the urban area, the profit of the producer in the local area decreases, and the profit of the producer in the urban area increases.

These results lead us to presume that the inefficiency due to the location in the local area decreases regional economic progress and accelerates the density of producers in the urban area. Notably, these results partially differ from the ordinary properties of a Bertrand equilibrium shown by Eaton and Grossman(1986) but are intuitively correct: the specific model of this paper indicates that an increase in the profit of one producer does not result in an increase in the profit of the other, despite the strategic complements under price competition, while the ordinal Bertrand case shows changes in profits in the same direction. Hence, the condition with respect to the shapes of iso-profit curves that enables government intervention to increase either of the profits should be studied further; the findings would contribute to the discussion on appropriate administrative guidance.

Moreover, using the specific model, we find that the profit of the producer in the local area always increases due to reinforcement of administrative guidance. We also show that when the inefficiency due to the location worsens, the effect of the administrative guidance that increases the profit in the local area shrinks, and subsequently, the local government decreases the level of administrative guidance to balance marginal profit and marginal cost for the local area. Hence, we conclude that when the inefficiency of the location worsens, for example, because of a drain of human resources to the urban area, the local government must eliminate this inefficiency to provide a sufficient effect of administrative guidance that corresponds to its cost, although it may be not as easy in practice as in theory; otherwise, the policy becomes ineffective compared with the cost incurred, and subsequently, the intervention of the local government can be extinguished in the long term.

Regarding resource allocation, because we suppose a third market model, the economic unit who can lose a part of their surplus because of an increase in food prices is the consumers in the foreign country. Thus, if we assume that the producers supply their food to the home market and consider the result of the second stage, we can surmise that a part of the surplus is transferred from home consumers to the producer in the local area when an improvement in the producer's differentiation strategies due to administrative guidance causes the price hike. Therefore, the local government should be considerate to balance the level of prices and resource allocation when setting administrative guidance if consumers are also in the home market.

One of the limitations of this study is the low robustness of the results in the second stage provided by the specific model (Appendix D). Although we suppose that the added value is constant for simplicity, further study of the endogenous added value would contribute to the field of Bertrand-fashioned strategic trade policy.

**Acknowledgements** Okimoto gratefully acknowledges Masanori Amano (University of Shizuoka), Tomoaki Imoto (University of Shizuoka) and the financial support of H29 Commendatory grants for research activities from University of Shizuoka.

## Appendix A : Stability Conditions

$k^U$ and $k^L$ are positive constants. For simplicity, define $y^U \equiv (p^{UE} - p^U)$ and $y^L \equiv (p^{LE} - p^L)$. Differentiating $Z$ in the text with respect to time, $\dot{Z} = k^U y^U (\dot{p}^{UE} - \dot{p}^U) + k^L y^L (\dot{p}^{LE} - \dot{p}^L)$ is given as the adjustment processes according to time. This formulation builds on the idea of a Liapunov function. If $\dot{Z} < 0$ when $Z > 0$ (the prices are not the equilibrium value) and $\dot{Z} = 0$ when $Z = 0$ (the prices are the equilibrium value), any prices converge to the equilibrium, and then the equilibrium is globally stable. Therefore, the condition which leads to $\dot{Z} < 0$ when $Z > 0$ is the stability condition.

### A.1. *Non-specific model*

Substituting the adjustment processes according to the reaction functions, $\dot{p}^{UE} = -\dfrac{a}{\frac{\partial X^U}{\partial p^U}+b}\dot{p}^L$ and $\dot{p}^{LE} = -\dfrac{c}{\frac{\partial X^L}{\partial p^L}+d}\dot{p}^U$, into $\dot{Z}$, we obtain

$$\dot{Z} = \dot{p}^U \left( -k^U y^U - k^L y^L \frac{c}{\frac{\partial X^L}{\partial p^L} + d} \right) + \dot{p}^L \left( -k^U y^U \frac{a}{\frac{\partial X^U}{\partial p^U} + b} - k^L y^L \right).$$

Furthermore, the adjustment processes according to time, $\dot{p}^U = k^U y^U$ and $\dot{p}^L = k^L y^L$, are substituted for the aforementioned function, which leads to

$$\dot{Z} = -(k^U y^U)^2 - (k^L y^L)^2 - k^U k^L y^U y^L \left( \frac{a}{\frac{\partial X^U}{\partial p^U} + b} + \frac{c}{\frac{\partial X^L}{\partial p^L} + d} \right).$$

Thus, the sign of $\dot{Z}$ depends on the signs of $y^U y^L$, $\dfrac{a}{\frac{\partial X^U}{\partial p^U}+b}$ and $\dfrac{c}{\frac{\partial X^L}{\partial p^L}+d}$.

We can show that if $0 < a < -b$ and $0 < c < -d$, $\dot{Z} < 0$ is satisfied. First, $0 < a < -b$ ensures $\dfrac{a}{\frac{\partial X^U}{\partial p^U}+b} < 0$ and $\left|\dfrac{a}{\frac{\partial X^U}{\partial p^U}+b}\right| < 1$. Hence, $0 < -\dfrac{a}{\frac{\partial X^U}{\partial p^U}+b} < 1$ holds. In the same manner, $0 < c < -d$ ensures $\dfrac{c}{\frac{\partial X^L}{\partial p^L}+d} < 0$ and $\left|\dfrac{c}{\frac{\partial X^L}{\partial p^L}+d}\right| < 1$; thus, $0 < -\dfrac{c}{\frac{\partial X^L}{\partial p^L}+d} < 1$ is obtained. These conditions can be then summarised as $\dfrac{a}{\frac{\partial X^U}{\partial p^U}+b} + \dfrac{c}{\frac{\partial X^L}{\partial p^L}+d} < 0$. Taking this into consideration, we find that $\dot{Z} < 0$ is clearly ensured when $y^U y^L < 0$ and $0 < -\dfrac{a}{\frac{\partial X^U}{\partial p^U}+b} - \dfrac{c}{\frac{\partial X^L}{\partial p^L}+d} < 2$ leads to

$$\dot{Z} = -(k^U y^U)^2 - (k^L y^L)^2 + k^U k^L y^U y^L \left( -\frac{a}{\frac{\partial X^U}{\partial p^U} + b} - \frac{c}{\frac{\partial X^L}{\partial p^L} + d} \right)$$

$$< -(k^U y^U)^2 - (k^L y^L)^2 + k^U k^L y^U y^L 2 = -(k^U y^U - k^L y^L)^2 \leq 0,$$

when $y^U y^L > 0$. Although the conditions that $0 > a > b$ and $0 > c > d$ enable $\dot{Z} < 0$ to hold, these conditions contradict strategic complements in the Bertrand model.

A.2. *Specific model*

The adjustment process according to the reaction functions can be represented as $\dot{p}^{UE} = \frac{1}{2}\dot{p}^L$ and $\dot{p}^{LE} = \frac{1}{2}\dot{p}^U$. Substituting those functions and the adjustment processes according to time, $\dot{p}^U = k^U y^U$ and $\dot{p}^L = k^L y^L$, into $\dot{Z}$, we obtain

$$\dot{Z} = k^U y^U \left( \frac{1}{2}\dot{p}^L - \dot{p}^U \right) + k^L y^L \left( \frac{1}{2}\dot{p}^U - \dot{p}^L \right) = -(k^U y^U)^2 - (k^L y^L)^2 + (k^U y^U)(k^L y^L).$$

Therefore, if $y^U y^L < 0$, $\dot{Z} < 0$ is ensured and if $y^U y^L > 0$, $\dot{Z} < -(k^U y^U - k^L y^L)^2 \leq 0$ is ensured, the equilibrium in the specific model is always stable, from the viewpoint of the adjustment process according to time.

# Appendix B : Slopes of iso-profit curves

In the non-specific model, a change in the slope of the iso-profit curve of producer U is

$$\frac{d}{dp^U}\left(\frac{dp^L}{dp^U}\right) = \frac{\partial}{\partial p^U}\left(\frac{dp^L}{dp^U}\right) + \frac{\partial}{\partial p^L}\left(\frac{dp^L}{dp^U}\right)\frac{dp^L}{dp^U}$$

$$= \frac{\partial}{\partial p^U}\left(\frac{dp^L}{dp^U}\right) + \frac{\partial}{\partial p^L}\left(\frac{dp^L}{dp^U}\right)\left[ -\frac{X^U + (p^U - c^U)\frac{\partial X^U}{\partial p^U}}{(p^U - c^U)\frac{\partial X^U}{\partial p^L}} \right]$$

$$= \frac{1}{(p^U - c^U)^2 \left(\frac{\partial X^U}{\partial p^L}\right)^3} \left\{ -\left(\frac{\partial X^U}{\partial p^U} + b\right)(p^U - c^U)\left(\frac{\partial X^U}{\partial p^L}\right)^2 \right.$$

$$\left. + 2a\frac{\partial X^U}{\partial p^L}\left[X^U + (p^U - c^U)\frac{\partial X^U}{\partial p^U}\right] - \left[X^U + (p^U - c^U)\frac{\partial X^U}{\partial p^U}\right]^2 \frac{\partial^2 X^U}{\partial p^{L2}} \right\}.$$

A change in that of producer L is

$$\frac{d}{dp^L}\left(\frac{dp^U}{dp^L}\right) = \frac{\partial}{\partial p^L}\left(\frac{dp^U}{dp^L}\right) + \frac{\partial}{\partial p^U}\left(\frac{dp^U}{dp^L}\right)\frac{dp^U}{dp^L}$$

$$= \frac{\partial}{\partial p^L}\left(\frac{dp^U}{dp^L}\right) + \frac{\partial}{\partial p^U}\left(\frac{dp^U}{dp^L}\right)\left[-\frac{X^L + (p^L - c^L)\frac{\partial X^L}{\partial p^L}}{(p^L - c^L)\frac{\partial X^L}{\partial p^U}}\right]$$

$$= \frac{1}{(p^L - c^L)^2 \left(\frac{\partial X^L}{\partial p^U}\right)^3}\left\{-\left(\frac{\partial X^L}{\partial p^L} + d\right)(p^L - c^L)\left(\frac{\partial X^L}{\partial p^U}\right)^2\right.$$

$$\left.+ 2c\frac{\partial X^L}{\partial p^U}\left[X^L + (p^L - c^L)\frac{\partial X^L}{\partial p^L}\right] - \left[X^L + (p^L - c^L)\frac{\partial X^L}{\partial p^L}\right]^2 \frac{\partial^2 X^L}{\partial p^{U^2}}\right\}.$$

With respect to those changes in the slopes, the sign of the denominator is positive, the sign of the first term in the numerator is also positive, and the signs of second and third terms in the numerator are indeterminate. Therefore, we naturally assume that $\frac{d^2 p^L}{dp^{U^2}} > 0$ and $\frac{d^2 p^U}{dp^{L^2}} > 0$, as in the text.

## Appendix C : Comparative statics

The results with respect to $t = R, G$ are

$$|J|\frac{dp^{UE}}{dt} = -\left[\frac{\partial^2 X^U}{\partial p^U \partial t}(p^U - c^U) + \frac{\partial X^U}{\partial t}\right]\left(d + \frac{\partial X^L}{\partial p^L}\right)$$

$$+ a\left[\frac{\partial^2 X^L}{\partial p^L \partial t}(p^L - c^L) - \frac{\partial X^L}{\partial p^L}\frac{\partial c^L}{\partial t} + \frac{\partial X^L}{\partial t}\right],$$

$$|J|\frac{dp^{LE}}{dt} = -\left(b + \frac{\partial X^U}{\partial p^U}\right)\left[\frac{\partial^2 X^L}{\partial p^L \partial t}(p^L - c^L) - \frac{\partial X^L}{\partial p^L}\frac{\partial c^L}{\partial t} + \frac{\partial X^L}{\partial t}\right]$$

$$+ \left[\frac{\partial^2 X^U}{\partial p^U \partial t}(p^U - c^U) + \frac{\partial X^U}{\partial t}\right]c.$$

Strictly, $\frac{dp^{jE}}{dt}$, $j = U, L$ should be denoted by $\frac{\partial p^{jE}}{\partial t}$.

## Appendix D: Robustness

We add the assumption that producer L chooses $R$ to maximise its profit, to our specific model:

$$\max_{p^U} \pi^U = (p^U - \bar{c})\frac{p^L - p^U}{P(G)R},$$

$$\max_{p^L, R} \pi^L = \left[p^L - \left(\bar{c} + \bar{c}^L + \alpha(R, G)\right)\right]\left(1 - \frac{p^L - p^U}{P(G)R}\right).$$

These maximisation problems lead to the following reaction functions:

$$\frac{\partial \pi^U}{\partial p^U} = 0 \Leftrightarrow p^U = \frac{p^L + \bar{c}}{2}, \tag{D.1}$$

$$\frac{\partial \pi^L}{\partial p^L} = 0 \Leftrightarrow p^L = \frac{1}{2}\left[P(G)R + p^U + \left(\bar{c} + \bar{c}^L + \alpha(R,G)\right)\right], \tag{D.2}$$

$$\frac{\partial \pi^L}{\partial R} = 0 \Leftrightarrow \frac{p^L - \left(\bar{c} + \bar{c}^L + \alpha(R,G)\right)}{P(G)R}\frac{p^L - p^U}{R} = \frac{\partial \alpha}{\partial R}\left(1 - \frac{p^L - p^U}{P(G)R}\right). \tag{D.3}$$

We assume $\frac{\partial^2 \alpha}{\partial R^2} = 0$ for simplicity. $\frac{\partial^2 \pi^L}{\partial R^2} = -\frac{p^L - p^U}{P(G)R^2}\left[\frac{\partial \alpha}{\partial R} + \frac{2}{R}(p^L - c^L)\right] < 0$ is ensured. The sufficient condition for producer L's maximum is assumed. $\frac{\partial \pi^L}{\partial p^L} = 0 \Leftrightarrow \frac{p^L - c^L}{P(G)R} = \left(1 - \frac{p^L - p^U}{P(G)R}\right)$ implies that Eq.(D.3) reduces to $\frac{p^L - p^U}{R} = \frac{\partial \alpha}{\partial R}$. Thus, we define $\frac{p^L - p^U}{R} = \frac{\partial \alpha}{\partial R}$ as Eq.(D.3)′. We totally differentiate Eqs.(D.1) − (D.2) and Eq.(D.3)′ to provide the properties of the equilibrium as

$$\begin{bmatrix} -2 & 1 & 0 \\ 1 & -2 & P(G) + \frac{\partial \alpha}{\partial R} \\ -1 & 1 & -\frac{\partial \alpha}{\partial R} \end{bmatrix} d\begin{bmatrix} p^U \\ p^L \\ R \end{bmatrix} = \begin{bmatrix} 0 \\ -1 \\ 0 \end{bmatrix} d\bar{c}^L + \begin{bmatrix} 0 \\ -\left(P'(G)R + \frac{\partial \alpha}{\partial G}\right) \\ \frac{\partial^2 \alpha}{\partial R \partial G} R \end{bmatrix} dG.$$

We define $|J|$ as the Jacobian matrix of the equation. Using Eq.(D.3)′, $|J|$ is equal to $P(G) - 2\frac{\partial \alpha}{\partial R}$. Thus, the results of the comparative statics, $\frac{\partial p^{UE}}{\partial \bar{c}^L} = \frac{-\frac{\partial \alpha}{\partial R}}{|J|}$, $\frac{\partial p^{LE}}{\partial \bar{c}^L} = \frac{-2\frac{\partial \alpha}{\partial R}}{|J|}$, and $\frac{\partial R^E}{\partial \bar{c}^L} = \frac{-1}{|J|}$, reduce to $\frac{\partial p^{UE}}{\partial \bar{c}^L} > 0$, $\frac{\partial p^{LE}}{\partial \bar{c}^L} > 0$, and $\frac{\partial R^E}{\partial \bar{c}^L} > 0$ (respectively, $\frac{\partial p^{UE}}{\partial \bar{c}^L} < 0$, $\frac{\partial p^{LE}}{\partial \bar{c}^L} < 0$ and $\frac{\partial R^E}{\partial \bar{c}^L} < 0$) if $P(G) < 2\frac{\partial \alpha}{\partial R}$ (respectively, $P(G) > 2\frac{\partial \alpha}{\partial R}$).

The negative indirect effect, $\frac{dR}{dp^U} < 0$, is considered the cause of indeterminacy of the signs of $\frac{\partial p^{UE}}{\partial \bar{c}^L}$ and $\frac{\partial p^{LE}}{\partial \bar{c}^L}$. Eq. (D.3)′ clarifies that $\frac{\partial X^{UE}}{\partial \bar{c}^L} = \frac{1}{P(G)R^E |J|}\left(-\frac{\partial \alpha}{\partial R} + \frac{p^{LE} - p^{UE}}{R^E}\right) = 0$ and $\frac{\partial X^{LE}}{\partial \bar{c}^L} = \frac{1}{P(G)R^E |J|}\left(\frac{\partial \alpha}{\partial R} - \frac{p^{LE} - p^{UE}}{R^E}\right) = 0$. Those are different from Proposition 1 and the second stage equilibrium in the text. Moreover, the amount of information on how the equilibrium prices are affected by $G$ is insufficient.